\documentstyle[prd,aps,amsmath,epsf,graphicx]{revtex}

\DeclareFontFamily{OT1}{rsfs10}{}
\DeclareFontShape{OT1}{rsfs10}{m}{n}{ <-> rsfs10 }{}
\DeclareMathAlphabet{\mathscript}{OT1}{rsfs10}{m}{n}

\newcommand{\be}{\begin{equation}}
\newcommand{\ee}{\end{equation}}

\newcommand{\bea}{\begin{eqnarray}}
\newcommand{\eea}{\end{eqnarray}}
\newcommand{\ba}{\begin{array}}
\newcommand{\ea}{\end{array}}

\newcommand{\tr}{\textrm{tr}}
\newcommand{\ns}{\normalsize}

\def\a{\alpha}
\def\b{\beta}
\def\g{\gamma}
\def\c{\chi}
\def\d{\delta}
\def\e{\epsilon}

\begin{document}

\begin{titlepage}

\title{
\hfill{\ns DCPT-05/27\\}
\hfill{\ns EMPG-05-12\\}
\hfill{\ns hep-th/0506092\\[2cm]}
{\Large On T-folds, G-structures and Supersymmetry.}\\[1cm]}
\setcounter{footnote}{0}
\author{{\ns\large
James Gray$^1$\footnote{email: j.a.gray2@durham.ac.uk} and
Emily Hackett-Jones$^2$\footnote{email: E.Hackett-Jones@ed.ac.uk }\\[1cm]}
{\ns $^1$Centre for Particle Theory }\\
{\ns Department of Mathematical Sciences} \\
{\ns University of Durham, Durham DH1 3LE, U.K.}\\[0.3cm]
{\ns $^2$School of Mathematics}\\ 
{\ns University of Edinburgh, Edinburgh EH9 3JZ, U.K.}}

\maketitle

\vspace{2cm}

\begin{abstract}
We describe how to calculate the amount of supersymmetry associated to
a class of supergravity theories 
obtained by compactification on T-folds. 
We illustrate our discussion by calculating 
the degree of supersymmetry enjoyed by a particular set of massive supergravities 
which have
been obtained in the literature by compactifying type II supergravity 
on such backgrounds. Our discussion involves a modification of the usual 
arguments, 
based upon G-structures, for the amount of supersymmetry preserved by geometric 
compactifications.
\end{abstract}

\thispagestyle{empty}

\end{titlepage}

\renewcommand{\thefootnote}{\arabic{footnote}}

%
\section{Introduction}
\label{intro}

The usual procedure for constructing background 
field configurations of string and M theory involves two steps. Firstly, the 
low energy fields are defined on coordinate patches in an atlas covering some manifold. A 
consistent global picture is then created in the second step by
insisting that where coordinate patches overlap the different field
configurations describe the same physical setup. 
This is usually achieved by relating the relevant local fields on different coordinate patches 
by diffeomorphisms
and gauge transformations. We shall refer to such backgrounds as ``geometric" as the resulting
configurations include a globally well defined metric tensor.

One can carry out the second step of requiring a consistent global picture 
in other ways (see \cite{Hull:2004in} for the first general consideration and discussion of this). 
Any symmetry of the theory can be used to join together coordinate patches
 while still fulfilling the criterion that the various field configurations on overlapping 
patches describe the same physics. For example, in a toroidally 
compactified type II string theory one could use elements of the
$O(d,d)$ T-duality group as transition functions
\cite{Hull:2004in,Greene:1989ya,Vafa:1996xn,Kumar:1996zx,Hellerman:2002ax,Dabholkar:2002sy,Kachru:2002sk,Hull:2003kr,Flournoy:2004vn,Lindstrom:2004iw,Hull:2005hk,Dabholkar:2005ve,Shelton:2005cf,Lawrence:2006ma,Schulz:2004ub,Flournoy:2005xe}. 
This results in 
what Hull has called a ``T-fold" \cite{Hull:2004in}. 
In general, there is no globally defined metric tensor on a T-fold. This is a consequence of the fact that T-duality 
transformations 
mix up the internal components of metric with those of the NS-NS two form. 
We will therefore follow the literature in referring to such spaces where the metric ``jumps'' 
between coordinate patches in this 
manner as ``non-geometric". The idea of patching a manifold together with duality transformations,
to obtain these so-called ``duality-folds'' \cite{Hull:2004in}, 
has been pursued by a number of authors over the last few years (see
\cite{Hull:2004in,Greene:1989ya,Vafa:1996xn,Kumar:1996zx,Hellerman:2002ax,Dabholkar:2002sy,Kachru:2002sk,Hull:2003kr,Flournoy:2004vn,Lindstrom:2004iw,Hull:2005hk,Dabholkar:2005ve} for this and related work). 
In passing we note that it appears that the use of supersymmetry in performing the `second step' above has 
not yet been considered in the literature.

Given their unusual nature it is helpful to briefly review some of the evidence that
string theory can make sense on such backgrounds. Firstly some T-folds can be shown to be 
T-dual to various, more standard, geometric spaces. For example, the T-dual of certain torus 
compactifications with NS-NS flux (with an appropriate domain wall solution in the external 
space) are T-folds (see for example \cite{Hellerman:2002ax,Kachru:2002sk}). Thus one is left with two 
possibilities. Either the T-fold obtained in such a manner is as good a string background as 
the torus compactification with flux, or T-duality is not a feature of the full theory in that, 
for some reason, the torus compactification's T-dual is not a good string background.

A second piece of evidence which indicates that we should take T-fold backgrounds seriously within the 
context of string theory is provided by their relation to certain well defined conformal 
field theories (CFTs) \cite{Hellerman:2002ax}. It is well known that certain smooth 
geometric backgrounds of string theory have limits in which 
they become orbifolds of flat space. The associated 
world sheet CFT can then be described in detail. In a similar manner there 
are limits of certain T-fold compactifications which are associated with well understood consistent
CFTs \cite{Hellerman:2002ax}. In this case the relevant orbifolds are {\it asymmetric} examples
in which the left and 
right moving degrees of freedom on the string world sheet are acted on differently by the 
orbifolding \cite{Narain:1986qm}.

Once it is accepted that it makes sense to consider string theory on these 
T-fold backgrounds 
the next obvious question is why are these configurations interesting? There are a multitude of 
interesting consequences which follow from considering string and M-theory on various symmetry
folds. One strong motivation for studying such backgrounds
is that they have been found to typically possess fewer unstabilised  moduli than more conventional
compactifications \cite{Hellerman:2002ax,Dabholkar:2002sy,Hull:2003kr,Flournoy:2004vn}.
It is reasonably straight forward to see why this is so.
Unstabilised moduli in lower dimensional theories correspond to unspecified integration constants in the 
associated higher dimensional vacuum solutions. The use of a larger set of symmetries (such as the full 
T-duality group instead of just some ``geometric'' subgroup) in patching together a compactification 
manifold effectively constitutes a more restrictive set of boundary conditions on the possible vacuum 
solutions. As such the solutions tend to have fewer unspecified integration constants, leading to fewer 
unstabilised moduli in the lower dimensional theories. However, within the context of phenomenology, where
considerations of moduli stabilisation would be most important, the use of these spaces raises 
various problems. This is because, for all values of the moduli, the known examples of non-trivial 
T-folds contain cycles of string scale size. This 
leads to concerns about the lack of a mass gap above the standard supergravity modes as well as 
concern that non-perturbative effects could well be important. These difficulties may well be 
resolvable in some cases. For example, it is easy to build examples of T-folds where one can 
show that, despite these comments, the size of non-perturbative effects in the superpotential is controlled by the 
volume of the base manifold and not that of the fibre \cite{wittenadiab}. In addition, there are such examples 
where only a finite number of extra light states would have to be included before a mass 
gap with the rest of the spectrum is obtained. 
However, in the interests of being conservative, and to show that our work is of interest whether
or not these issues are resolvable, we shall state our motivation in this paper in terms of a 
different use for T-fold compactification. 

An uncontroversial use of T-fold compactification is to regard it as a formal tool for finding 
new massive supergravity theories. Compactification of massless higher dimensional 
type II supergravities on T-folds results in examples of lower dimensional
massive supergravities, which in some cases were previously unknown \cite{Dabholkar:2002sy,Hull:2003kr,Hull:2005hk}. 
The supergravities obtained in this manner can be thought of as the 
completion of the class of such theories obtained by compactification on geometric spaces 
(for examples based upon Scherk Schwarz reduction see 
\cite{Scherk:1979zr,Scherk:1978ta,Salam:1984ft,Bergshoeff:1996ui,Lavrinenko:1997qa,Kaloper:1998kr,Hull:1998vy,Hull:2002wg,Bergshoeff:1997mg,Meessen:1998qm,Cowdall:1996tw,Cowdall:2000sq,Bergshoeff:2002mb,Kaloper:1999yr,Cvetic:2003jy,Bergshoeff:2003ri,AlonsoAlberca:2003jq,deWit:2002vt}).
In particular, these new sugras fit nicely into certain classifications of 
such theories, one of which is based on an $O(d,d)$ group of transformations \cite{Kaloper:1999yr,Hull:2005hk}.

One of the most basic questions one would like to answer about these lower dimensional supergravity 
theories is how much supersymmetry they possess. This reduces to a question about the nature of the truncation of the 
higher dimensional fields which is used in the reduction. 
More precisely, the question is which of the higher dimensional supersymmetries
are compatible with the  different forms of truncation that one might use?
This is the question we shall answer in part in this paper - how many
supersymmetries are enjoyed by a theory obtained by T-fold
compactification with a certain natural choice of truncation?

Our discussion will make it clear how to calculate the number of such supersymmetries associated with 
a large class of theories resulting from T-fold compactification. However, in the interests of clarity we will 
illuminate our considerations with the example of compactifications of type II string
theory on T-folds which take the form of a $T^d$ bundle over $S^1$, where the torus
experiences a monodromy in $SO(d,d,Z)$ around the base $S^1$. 
Similar compactifications have been considered in Refs.~\cite{Dabholkar:2002sy,Hull:2003kr,Hull:2005hk}.
We will extend this work by incorporating extra terms originating from 
the NS B-field in the dimensional reduction and will 
then proceed to calculate the amount of supersymmetry that is manifest in 
the compactified theories obtained.
To do this we will use and extend
work in the literature~\cite{Hassan:1999bv,Hassan:1999mm,Hassan:2000kr} where the rules
for the transformations of supersymmetry parameters under T-duality have been given. 
We will then use a modified version of the criterion for preservation of supersymmetry in 
dimensional reduction, phrased in terms of 
G-structures \cite{Gauntlett:2001ur,Gauntlett:2002sc,Friedrich:2001nh,Friedrich:2001yp,Ivanov:2001ma,Gurrieri:2002wz,Gauntlett:2002fz,Kaste:2002xs,Gurrieri:2002iw,Kaste:2003dh,Behrndt:2003uq,Gauntlett:2003cy,Gauntlett:2003wb,Behrndt:2003zg,Fidanza:2003zi,Frey:2004rn,Jeschek:2004je,Grana:2004bg,Gurrieri:2004dt,House:2004pm,Jeschek:2004wy,Tomasiello:2005bp,Behrndt:2005bv,House:2005yc,Grana:2005sn,Grana:2005ny,Haack:2001iz,Singh:2002tf,Singh:2001gt,LopesCardoso:2002hd,LopesCardoso:2003sp,Dall'Agata:2003ir,Dall'Agata:2004dk,LopesCardoso:2004ni}, to determine the number of supersymmetries
manifest in the theory. The approach we will follow, based
on the work of \cite{Hassan:1999bv,Hassan:1999mm,Hassan:2000kr}, will
lead us to consider the supersymmetry of theories which result
from compactification upon T-folds constructed using a 
certain minimal version of compactified supergravity. In
particular, we will not consider cases arising from 
constructions where a coset
reformulation of the theory obtained by reducing on the fibre of 
the T-fold is utilised in forming the overall compact space. See Appendix~A and later 
sections of the paper for more details.

The plan of this paper is as follows. In section II we shall introduce the T-fold 
backgrounds that we will be using as examples in the rest of the paper. We will then review
dimensional reduction on such a background to illustrate the kind of massive supergravity that 
can be obtained in this manner. In section III we shall describe how to calculate the
amount of supersymmetry associated with a class of theories obtained by 
T-fold reduction using the examples of the previous 
section to illustrate our method. We will describe first how the usual arguments proceed in 
the case of geometric dimensional reduction before going on to describe how this analysis 
changes in the T-fold case. In section IV we briefly conclude.

\section{Examples of Massive Supergravities from T-fold Reduction.}
\subsection{Examples of T-fold vacua for dimensional reduction.}\label{Tfolds}

The examples of duality-folds that we will consider are the class of T-folds which 
have been studied in \cite{Dabholkar:2002sy,Hull:2003kr}. We start
by considering a type II superstring theory on a $T^d$ bundle over 
$M_{9-d} \times S^1$ (the discussion and notation of this subsection follows that of \cite{Hull:1998vy}). 
However, we are not going to 
take the trivial vacuum on this space. Instead we consider a situation where the various fields 
in the theory have a dependence on the $S^1$,
which has coordinate $y \sim y + 2 \pi$, of the following form,
\bea
\label{ansatz}
\psi(y) = g(y)[\psi]
\eea
Here $\psi$ is a general field which is taken to be independent of the $T^d$ directions. We then make a 
Scherk-Schwarz ansatz \cite{Scherk:1979zr,Scherk:1978ta} for $g(y)$,
\bea
g(y)=\exp \left( \frac{y T}{2 \pi} \right)
\eea
where $T$ is in the Lie algebra of $SO(d,d)$, which is a subgroup of
the T-duality group associated with the $T^d$ fibre. 
This ansatz ensures that a solution to the lower dimensional theory,
obtained by compactification on this vacuum, is also a solution to 
the higher dimensional equations of motion up to the relevant approximations. 
As such it guarantees that we do not need to worry about such subtleties as compensators in the 
dimensional reduction.
The resulting field configurations are not periodic on traversing the $S^1$. Instead there 
is a monodromy,
\bea
{\cal M}(g) = e^T \;
\eea
This is simply an element of $SO(d,d)$. In other words, when we
traverse the circle once the fields are not identified in the usual
way, but instead come back to themselves up to
an $SO(d,d)$ element of the T-duality group. In fact we should choose the monodromy to 
be within $SO(d,d,Z)$ as this is the relevant symmetry subgroup of the full
theory when massive states, which
we have truncated, are included. Note that since we are
making this Scherk-Schwarz ansatz we will only be able to consider
elements of the full T-duality group, $O(d,d,Z)$, which are continuously connected
to the identity in $O(d,d,R)$.

The resulting vacuum is an example of a T-fold \cite{Hull:2004in}. 
The field configurations at $y=0$ and $y=2 \pi$ 
are not simply related by diffeomorphisms and gauge transformations. A 
less trivial element of the T-duality group is required to transform them into one another. 
This vacuum is then non-geometric 
in the sense that there is not a globally well defined metric in ten dimensions. 
This is because the transition functions mix up the metric and NS-NS two form on
the toroidal fibre when we go once around the circle.

\subsection{Massive supergravities from dimensional reduction on T-folds.}
\label{theory}

We will now describe how the low energy effective action associated with a 
dimensional reduction on one of the T-folds of the 
previous subsection is obtained. Many, but not all, of the terms in the reduced theory that we 
will present have been obtained elsewhere \cite{Dabholkar:2002sy,Hull:2003kr}. 
However, it is useful to provide a discussion of the 
dimensional reduction here both to keep this paper reasonably self contained and also to 
complete the dimensional reduction of the ten dimensional 
NS-NS sector. To our knowledge some terms of this reduction, 
descending from the three form field strength, have 
yet to appear in the literature. Since this paper first appeared the relation between reductions 
of the type we are going to present here and so called 'twisted torus' reductions (see for 
example \cite{Hull:2005hk,Kaloper:1999yr}) has been demonstrated in \cite{Dabholkar:2005ve}. 

For the sake of brevity we will
consider a $T^2$ fibre, i.e. we will compactify on a $T^2$ bundle over
$S^1$. The generalisation of this to the $T^d$ case is straight-forward.
The starting point for our dimensional reduction is the ten dimensional low energy effective 
action of type II superstring theory. Again,
for the sake of brevity, and because it is common to both IIA and IIB, we only consider 
the NS-NS sector.

\bea
S_{10} = \int_{{\cal M}^{10}} dx \sqrt{-g} e^{-\phi} \left[R +
  (\partial \phi)^2 - \frac{1}{12} H^2 \right]
\eea

The simplest procedure to follow for the reduction is to first reduce on the $T^2$ fibre and then perform a further reduction on 
the base space incorporating our non-trivial twist. As such we require an effective eight dimensional action for the 
reduction of the above theory on the $T^2$ fibre in which the $O(2,2)$ duality group is manifest. Such 
an eight dimensional action is given in the seminal paper by Maharana and Schwarz \cite{Maharana:1992my}. 
We will not repeat their calculation here but shall simply quote the results. They obtain,
\bea
S_{8D} = \int_{{\cal M}^8} d^8 x \sqrt{-g} e^{-\phi^{(8)}} \left[ R^{(8)} + (\partial \phi^{(8)})^2 + 
\frac{1}{8} \tr (\partial M^{-1}
\partial M) - \frac{1}{4} {\cal F}^i_{\mu \nu} M^{-1}_{ij} {\cal F}^{\mu \nu j} - \frac{1}{12} H^2 \right] \; \label{8daction},
\eea
where the three form field strength and eight dimensional dilaton are defined by,
\bea
\label{hdef}
H_{\mu \nu \rho} &=& \partial_{\mu} B_{\nu \rho} - \frac{1}{2} {\cal A}^i_{\mu} \eta_{i j} {\cal F}^i_{\nu \rho} + 
\textnormal{cyclic permutations} \;, \\
\phi^{(8)} &=& \phi - \frac{1}{2} \log \det G \;.
\eea
In these expression we have also defined,
\bea
M = \left(\ba{cc} G^{-1} & -G^{-1}B \\ B G^{-1} & G - B G^{-1} B \ea \right)\label{Mdef} \\
\eea
where $B$ and $G$ are the NS two form and metric on the fibre
respectively. Finally, if we denote the two coordinate indices on the fibre
as $\theta^i$, the gauge potentials are given by
\bea
{\cal A}^i_{\mu} &=& G_{\mu \theta^i} \;\;\; i=1,2, \\ \nonumber
{\cal A}^i_{\mu} &=& B_{\mu \theta^{i-2}} + B_{\theta^{i-2} \theta^j} {\cal A}_{\mu}^{j} \ \;\;\; i=3,4,
\eea
and the field strengths are simply ${\cal F}^i = d {\cal A}^i$.

In this formulation, the various fields transform under an $O(2,2)$ transformation as follows,
\bea
M &\rightarrow& \Omega M \Omega^T \label{Mtransf}\\
{\cal A}_{\mu} &\rightarrow& \Omega {\cal A}_{\mu}
\eea
Here $\Omega$ is the $O(2,2)$ matrix, satisfying $\Omega^T \eta \Omega =
\eta$, where in our conventions the invariant metric is given by,
\bea
\eta = \left(\ba{rr} 0 & 1 \\ 1 & 0 \ea \right) \;.
\eea
Given the $O(2,2)$ covariant form of the action in (\ref{8daction}), we now
make an ansatz for the dimensional reduction on the base of our
T-fold. From the discussion in the previous subsection our ansatz takes the 
following form,
\bea
ds_8^2 &=& g^7_{\alpha \beta}(x^{\gamma}) dx^{\alpha} dx^{\beta} + e^{2 \alpha} \left( dy + A^B_{\alpha}(x^{\gamma}) 
dx^{\alpha} \right)^2 \\
\phi^{(8)} &=& \phi^{(8)}(x^{\alpha}) \\
\label{ydepM}
M &=& \Omega(y) M_0(x^{\alpha}) \Omega^T(y) \\
{\cal A}^i &=& \Omega^i_j(y) {\cal A}^{0 j}_{\alpha}(x^{\beta}) dx^{\alpha} + 
\Omega^i_j(y) {\cal A}^{0 j}_{y}(x^{\beta}) dy \\
B &=& \frac{1}{2} B_{\alpha \beta}(x^{\gamma}) dx^{\alpha} \wedge dx^{\beta} + B^B_{\alpha}(x^{\gamma}) dx^{\alpha} 
\wedge dy
\eea
where  $x^{\alpha}$ are coordinates on the 7-dimensional non-compact space and
\bea
\Omega(y) = e^{\frac{y}{2 \pi} T}
\eea
defines the monodromy. We can now perform the reduction on the base space. 

\vspace{0.2cm}

The reduction of the Ricci scalar and dilaton terms to seven dimensions is unaffected by our duality twists. As such
the expressions for these terms are the standard ones.

\bea
S_1 = \int_{{\cal M}^7} dx \sqrt{-g^7} e^{-\phi^{(7)}} \left[R^{(7)} - (\partial \alpha )^2  + (\partial \phi^{(7)})^2
 - \frac{1}{4} e^{2 \alpha} (F^{B})^2\right] 
\eea
Here we have defined the usual seven dimensional dilaton $\phi^{(7)} = \phi^{(8)} - \alpha$.

The reduction of the remaining scalar kinetic terms does get modified by the duality twist. The resulting kinetic terms 
are the usual ones but the derivative involved is replaced by a covariant derivative with a non-trivial connection. In 
addition a potential for these fields is obtained. This potential and its properties have been discussed in some detail in
\cite{Dabholkar:2002sy}. The relevant terms are,

\bea
S_2 = \int_{{\cal M}^7} dx \sqrt{-g^7} e^{-\phi^{(7)}} \left[\frac{1}{8} \textnormal{tr} \left( D M_0^{-1} D M_0 \right) -
\frac{1}{4 (2 \pi)^2} \textnormal{tr}( T^T M_0^{-1} T M_0 + T^2)  \right]   \;.\label{scalarpotential}
\eea

Here we have defined a covariant derivative of $M_0$ as follows.

\bea
D_{\alpha} M_0 = \partial_{\alpha} M_0 - \frac{T M_0}{2 \pi} A^B_{\alpha} - \frac{M_0 T^T}{2 \pi} A^B_{\alpha}
\eea
where $M_0(x^{\alpha})$ is defined by $M(x^{\alpha}, y) =
\Omega(y) M_0(x^{\alpha}) \Omega(y)^T$.

\vspace{0.2cm}

The reduction of the vector field strength terms is a little more subtle. If we wish the result of the dimensional 
reduction to take an elegant form then we must ensure that we have made reasonable choices 
for our definitions of the seven dimensional fields \cite{Maharana:1992my,Kaloper:1999yr}.

For example, consider the behaviour of the vector fields under an $x^{\alpha}$ dependent shift of the $y$ coordinate,
$dy \rightarrow dy^{\prime} = dy + \partial_{\alpha} \omega^y dx^{\alpha}$.

\bea
{\cal A} \rightarrow {\cal A}^{\prime} = {\cal A}^{\prime 0 i}_{\alpha} dx^{\alpha} + {\cal A}^{\prime 0 i}_{y} \left(
dy + \partial_{\alpha} \omega^y dx^{\alpha} \right)
\eea

Thus we find that, while ${\cal A}^{0 i}_{y}$ is unchanged by such a transformation, ${\cal A}^{0 i}_{\alpha}$ is 
not a gauge invariant definition of a seven dimensional field.

\bea
{\cal A}^{0 i}_{\alpha} \rightarrow 
{\cal A}^{\prime 0 i}_{\alpha} = {\cal A}^{0 i}_{\alpha} - \partial_{\alpha} \omega^y {\cal A}^{ 0 i}_{y}
\eea

The only seven dimensional gauge field we wish to transform non-trivially under a shift of the base coordinate is the 
associated Kaluza Klein gauge field $A^B$. As such, instead of using ${\cal A}^{0 i}_{\alpha}$ as a seven dimensional 
field we make a field redefinition to obtain a quantity that has the properties we desire.
 
\bea
{\cal A}^{ 7 i}_{\alpha} = {\cal A}^{0 i}_{\alpha} - { A}^{B}_{\alpha} {\cal A}^{0 i}_{y}
\eea

Having decided on a set of definitions for the seven dimensional fields we may proceed with the dimensional reduction 
of the vector field strength terms. These terms are affected by the duality twists and so we obtain 
modifications to the result obtained in an untwisted reduction which depend on the generator $T$. We find for the 
components of the field strength,

\bea
{\cal F}^j_{\alpha \beta} &=& \Omega^j_i ({\cal F}^{7 i}_{\alpha \beta} + 2 A^B_{[\beta} \partial_{\alpha]} {\cal A}^{0 i}_y 
+  {\cal A}^{0 i}_y F^B_{\alpha \beta} )\\
{\cal F}^j_{\alpha y} &=&  \Omega^j_i ( \partial_{\alpha} {\cal A}^{0 i}_y  - \frac{T}{2 \pi} A^B_{\alpha} {\cal A}^{0 i}_y 
- \frac{T}{2 \pi} {\cal A}^{7 i}_{\alpha} )\;.
\eea

Here we have defined a field strength for ${\cal A}^{7 i}_{\alpha}$ in the usual way. As always in such reductions the 
easiest way to proceed is to switch to using an orthonormal basis during the reduction and then return to a coordinate 
basis once the result has been obtained. We define the following vielbeins (barred indices denote those in an orthonormal
basis).

\bea
e^{\mu}_{\bar{\nu}} = \left(\ba{rr} e^{7 \alpha}_{\bar{\alpha}}& - e^{7 \beta}_{\bar{\alpha}} A^B_{\beta}\\ 
0 & e^{-\alpha}\ea \right)\;\;\; e^{\bar{\mu}}_{\nu} = \left( \ba{rr} 
e^{7 \bar{\alpha}}_{\alpha} & e^{\alpha} A^B_{\alpha} \\
0 & e^{\alpha} \ea \right)
\eea

In these expressions, $e^{7 \bar{\gamma}}_{\alpha} e^{7 \bar{\gamma}}_{\beta} = g^7_{\alpha \beta}$, etc. in the usual 
manner. Using these definitions we then find the following.

\bea
-\frac{1}{4} {\cal F}^i M^{-1}_{ij} {\cal F}^j = -\frac{1}{4} {\cal F}^{i}_{\bar{\alpha} \bar{\beta}} M_{ij}^{-1} {\cal F}^{j \bar{\alpha} \bar{\beta}
} - \frac{1}{2} {\cal F}^{i}_{\bar{\alpha} \bar{y}} M_{ij}^{-1} {\cal F}^{j \bar{\alpha} \bar{y}}
\eea
where,
\bea
{\cal F}^i_{\bar{\alpha} \bar{\beta}} &=& e^{7 \alpha}_{\bar{\a}} e^{7 \b}_{\bar{\b}} {\cal F}^i_{\a \b} - 
2 e^{7 \alpha}_{[\bar{\a}} e^{7 \g}_{\bar{\b}]} A^B_{\g} {\cal F}^i_{\a y} \\
{\cal F}^i_{\bar{\a} \bar{y}} &=& e^{7 \alpha}_{\bar{\a}} e^{-\a} {\cal F}^i_{\a y} \; .
\eea

After a little algebra the following result is obtained for the reduction of the vector field strength terms.

\bea
S_3 = \int_{{\cal M}^7} dx \sqrt{-g^7} e^{-\phi^{(7)}} \left[
  -\frac{1}{4} \tilde{\cal F}^i_{\a \b} M^{-1}_{0ij} 
\tilde{\cal F}^{ j \a \b} - \frac{1}{2} e^{-2 \a} D_{\a} {\cal A}^{0 i}_y M_0^{-1 ij} D^{\a} {\cal A}^{0 j}_y \right]
\eea
Here we have used the following definitions of field strengths and covariant derivatives.

\bea
\tilde{\cal F}^i_{\a \b} &=& \tilde{\cal F}^{7 i}_{\a \b} + {\cal A}^{0 i}_y F^B_{\a \b} \\
\tilde{\cal F}^{7 i}_{\a \b} &=& {\cal F}^{7 i}_{\a \b} + 2 A^B_{[ \beta} \frac{T^i_j}{2 \pi} {\cal A}^{7 j}_{\a ]} \\
D_{\a} {\cal A}^{0 i}_y &=& 
\partial_{\a} {\cal A}^{0 i}_y - \frac{T^i_j}{2 \pi} A^B_{\a} {\cal A}^{0 j}_y - \frac{T^i_j}{ 
2 \pi} {\cal A}^{7 j}_{\a}
\eea

\vspace{0.2cm}

This just leaves us with the dimensional reduction of the anti-symmetric tensor terms to perform. Despite the fact that 
the field strength $H$ is itself invariant under the duality transformations we are using, the matrix $T$ still enters 
this part of the reduction. The reason for this is that the field strength, while invariant overall, is made up of 
components which transform non trivially as can be seen from its definition \eqref{hdef}.

As was the case for the reduction of the vector field part of the action we need to be careful to choose `gauge 
invariant' definitions of our seven dimensional degrees of freedom. We find that $B^B_{\a}$ has the properties we 
desire but that $B_{\a \b}$ does not. Therefore we define a new two form potential as follows.

\bea
B^7_{\a \b} = B_{\a \b} + A^B_{[ \a} B^B_{\b ]}
\eea

Then we can proceed with the dimensional reduction of these terms making use of our orthonormal basis 
as we did for the vector field strength piece. We find,

\bea
-\frac{1}{12} H^2 = -\frac{1}{12} H_{\bar{\a} \bar{\b} \bar{\g}} H_{\bar{\a} \bar{\b} \bar{\g}} - 
\frac{1}{4} H_{\bar{\a} \bar{\b} \bar{y}} H_{\bar{\a} \bar{\b} \bar{y}}
\eea

where,

\bea
H_{\bar{\a} \bar{\b} \bar{\g}} &=& e^{7 \a}_{\bar{\a}} e^{7 \b}_{\bar{\b}} e^{7 \g}_{\bar{\g}} H_{\a \b \g} - 
(e^{7 \d}_{\bar{\a}} e^{7 \b}_{\bar{\b}} e^{7 \g}_{\bar{\g}} A^B_{\d} H_{y \b \g} + \textnormal{2 perms.} )\\
H_{\bar{\a} \bar{\b} \bar{y}} &=& e^{7 \a}_{\bar{\a}}e^{7 \b}_{\bar{\b}} e^{-\a} H_{\a \b \g}
\eea

and

\bea
H_{\a \b \g} &=& \partial_{\a} B^{7}_{\b \g} - \frac{1}{2} F^B_{\a \b} B^B_{\g} + \frac{1}{2} H^B_{\a \b} A^B_{\g}
- \frac{1}{2} ({\cal A}^{7 i}_{\a} + A^B_{\a} {\cal A}^{0 i}_y) \eta_{i j} ({\cal F}^{7 i}_{\b \g} + {\cal A}^{0 i}_y 
F^B_{\b \g} ) - ({\cal A}^{7 i}_{\a}+ A^B_{\a} {\cal A}^{0 i}_y)\eta_{i j} A^B_{[ \g } \partial_{\b ]} {\cal A}^{0 i}_y \\
H_{\a \b y} &=& H^B_{\a \b} - \frac{1}{2} {\cal A}^{0 i}_y \eta_{ij} ( {\cal F}^j_{\a \b} + 2 A^B_{[ \b} \partial_{\a ]}
{\cal A}^{0 i}_y + {\cal A}^{0 i}_y F^B_{\a \b} ) - ({\cal A}^{7 i}_{[\a} + A^B_{[ \a} {\cal A}^{0 i}_{|y|}) \eta_{ij} 
D_{\b ]} {\cal A}^{0 i}_y\; .
\eea

In the above $H^B$ is the field strength associated with $B^B$ defined in the usual manner. 
After a little algebra we then find the following for the final piece of our seven dimensional action.

\bea
S_4 = \int_{{\cal M}^7} dx \sqrt{g^7} e^{-\phi^{(7)}} \left[ - \frac{1}{12} (H^7)^2 - \frac{1}{4} e^{-2 \a} 
(H^B_{\a \b} - \frac{1}{2} {\cal A}^{0 i}_y \eta_{ij} \tilde{{\cal F}}^j_{\a \b} - {\cal A}^{7 i}_{[ \a} \eta_{ij} D_{\b ]} {\cal A}^{0 i}_y 
)^2 \right]
\eea
In this expression we have used the following definition of the seven dimensional three-form field strength.

\bea
H^7_{\a \b \g} = \partial_{\a} B^7_{\b \g} - \frac{1}{2} F^B_{\a \b} B^B_{\g} - \frac{1}{2} H^B_{\a \b} A^B_{\g} - \frac{1}{2}
{\cal A}^{7 i}_{\a} \eta_{ij} \tilde{{\cal F}}^j_{\b \g} + \textnormal{ cyclic perms.}
\eea

\vspace{1cm}

Combining all the pieces we have obtained in this subsection we then obtain the following for the reduction of 
the NS-NS sector of type II string theory on this class of T-folds.

\bea
S_7 = \int_{{\cal M}^7} \sqrt{-g^7} e^{-\phi^{(7)}} \left[R^{(7)} - (\partial \alpha )^2  + (\partial \phi^{(7)})^2
 - \frac{1}{4} e^{2 \alpha} (F^{B})^2 \right. \\ \nonumber \left. +\frac{1}{8} \textnormal{tr} \left( D M_0^{-1} D M_0 \right) -
\frac{1}{4 (2 \pi)^2} \textnormal{tr}( T^T M_0^{-1} T M_0 + T^2)
\right. \\ \nonumber \left. -\frac{1}{4} \tilde{\cal F}^i_{\a \b} M^{-1}_{0ij} 
\tilde{\cal F}^{ j \a \b} - \frac{1}{2} e^{-2 \a} D_{\a} {\cal A}^{0 i}_y M_0^{-1 ij} D^{\a} {\cal A}^{0 j}_y 
\right. \\ \nonumber \left. - \frac{1}{12} (H^7)^2 - \frac{1}{4} e^{-2 \a} 
(H^B_{\a \b} - \frac{1}{2} {\cal A}^{0 i}_y \eta_{ij} \tilde{{\cal F}}^j_{\a \b} - 
{\cal A}^{7 i}_{[ \a} \eta_{ij} D_{\b ]} {\cal A}^{0 i}_y 
)^2 \right]
\eea

In this action we have the following field strength and covariant derivative definitions.

\bea
F^B_{\a \b} &=& 2 \partial_{[ \a} A^B_{ \b]} \\
H^B_{\a \b} &=& 2 \partial_{[ \a} B^B_{ \b]} \\
\tilde{{\cal F}}^i_{\a \b} &=& 2 \partial_{[\a} {\cal A}^{7 i}_{\b ]} + {\cal A}^{0 i}_y F^B_{\a \b} + 2 A^B_{[ \b} \frac{T^i_j}{2 \pi} 
{\cal A}^{7 j}_{\a ]} \\
H^7_{\a \b \g} &=& \partial_{\a} B^7_{\b \g} - \frac{1}{2} F^B_{\a \b} B^B_{\g} - \frac{1}{2} H^B_{\a \b} A^B_{\g} - \frac{1}{2}
{\cal A}^{7 i}_{\a} \eta_{ij} \tilde{{\cal F}}^j_{\b \g} + \textnormal{ cyclic perms.} \\
D_{\alpha} M_0 &=& \partial_{\alpha} M_0 - \frac{T M_0}{2 \pi} A^B_{\alpha} - \frac{M_0 T^T}{2 \pi} A^B_{\alpha} \\
D_{\a} {\cal A}^{0 i}_y &=& 
\partial_{\a} {\cal A}^{0 i}_y - \frac{T^i_j}{2 \pi} A^B_{\a} {\cal A}^{0 j}_y - \frac{T^i_j}{ 
2 \pi} {\cal A}^{7 j}_{\a}
\eea

It is easy to show that this action is that obtained by reducing the NS-NS part of the ten 
dimensional action on a three torus, supplemented by various mass and charge terms which are 
determined by the duality twist generator $T$. This is an example of the 
kind of massive supergravity we will consider in the next section. In
particular, we will calculate the amount of supersymmetry this type of
theory possesses.

In the low energy supergravity limit we have discussed here, the compactifications which have monodromies 
within the same $O(d,d,R)$ conjugacy class are equivalent and so give rise to the same reduced theory \cite{Dabholkar:2002sy}. This point will tie in nicely with the 
discussion of supersymmetry that will appear in the next section. It should also be pointed out that the above reduction
implicitly uses the fact that the usual rules for flux quantisation in string theory are modified when the compact cycles
being considered are nongeometric. There are scalar fields in the lower dimensional theory which originate from the components 
of the higher dimensional NS two form with indices on the compact space. For some of these fields, the fact that these 
degrees of freedom appear as lower dimensional fields which are able to vary in value depends crucially on the fact 
that the integral of the three-form field strength over certain cycles is not quantised in the usual manner.

\section{Supersymmetry and T-fold reduction}

This section is split into three main parts. Firstly we shall briefly review the standard discussion of 
how one determines the amount of supersymmetry associated with a
geometric dimensional reduction. 
In the second subsection we will describe the transformation of various quantities 
under T-duality. Finally, we will use this information to describe how the 
standard discussion for the degree of supersymmetry of a dimensionally reduced theory is 
modified in the T-fold case.

A geometrical 
dimensional reduction proceeds in a series of steps. First one chooses a Riemannian manifold 
$X$, with metric and ansatze for the other fields, on which to compactify the extra dimensions. 
One then proceeds to rewrite the fields in the higher dimensional theory in a manner that respects 
the symmetries of this background. The fields are split up into representations of the subgroup of 
the full symmetry group of the theory that is preserved by the background. 
Up until this point nothing in the theory has actually changed. One has simply relabeled various 
fields in such a way as to make the rest of the dimensional reduction easier. The change to the 
content of the theory comes from the rest of the process. The extra 
dimensional dependence of the fields is expanded in a series of modes about the background and 
various parts of the expanded and 
rewritten higher dimensional fields, such as some of these modes, are discarded.
The fields are then substituted into the higher 
dimensional action, compensators if needed are calculated, and the extra dimensions are integrated 
out to leave the lower dimensional theory. All of this has to be
performed in a consistent manner so that every 
solution to the resulting low energy theory can be associated to a solution of the higher 
dimensional one, up to any approximations that are made in the truncation - which {\it defines} the 
lower dimensional theory. Furthermore, one might wish the truncation to be `physical' in that
the modes one discards in this process should describe fluctuations which are more massive than 
those which are kept. The classic example of this is Kaluza Klein reduction on a torus.
In general, however, figuring out how physically to define the truncation is a highly non-trivial task.

Fortunately, for the Scherk-Schwarz reductions considered in the
previous section a consistent truncation is easily defined. In fact, the
ansatze for the fields already includes this truncation of the full 
possible field content of the theory.

The question we are interested in is how much supersymmetry is preserved by this sort of process 
(note we are interested in the amount of supersymmetry preserved by the lower dimensional {\it theory}
as opposed to any specific vacuum of it).
Supersymmetry, like any symmetry, 
 describes the fact that if one mixes up the various fields in the theory in a certain manner
then one gets back to the same action as one began with. Given the above discussion, the crucial 
question  
becomes whether this mixing up of the fields is consistent with the expansion and truncation that has 
been made of the fields - 
no other procedure in dimensional reduction changes the theory and so no other procedure can 
break supersymmetry. Therefore, a supersymmetry will be preserved in a dimensional reduction if this 
process is such that the supersymmetry transformations only mix up parts of the higher dimensional
fields which were kept in the truncated action with other parts that
were kept.  If this is not the case then the supersymmetry can not be preserved in
the lower dimensional theory.

For commonly used geometrical compactifications and truncations there
is a single mathematical rule which 
tells us which supersymmetries are preserved by dimensional 
reduction on direct product spacetimes. A supersymmetry is preserved if
its parameter, $\epsilon$, 
can be written as a product of an external and an internal spinor, the latter being a singlet under
the structure group of the spin bundle associated to the internal manifold 
\footnote{Since in a geometric compactification the internal manifold is Riemannian it has a 
metric tensor which constitutes an
$O(d)$ structure for a $d$ dimensional compact geometric space. 
The structure group of the frame bundle of the manifold is reduced from $Gl(d)$ to 
(a subgroup of) $O(d)$ in this case. Thus the frame bundle admits a principle sub-bundle which is 
the orthonormal
frame bundle. If the manifold is spin this principle bundle has an associated spin bundle - sections
of which form the spinor fields on the manifold.}. In other words, the 
internal part of the supersymmetry parameter should be globally defined and nowhere vanishing on the 
internal space in order for the associated supersymmetry to be preserved.

This rule is certainly true for the standard Kaluza Klein toroidal reduction where one truncates to zero 
modes - in this case all supersymmetry is then preserved. We shall
show shortly that this rule is also true for the kind 
of Scherk Schwarz reductions we considered in the previous section when the monodromy is taken to 
lie in a geometric subgroup of $O(d,d)$. It is true for truncation to the massless 
modes on a Calabi-Yau reduction. In more complicated cases the same
rule also holds; indeed in \cite{Grana:2005ny} a 
truncation was proposed for reductions on manifolds of $SU(3)$ structure with non-vanishing intrinsic 
torsion which was constructed to obey this rule.

In the general case where the structure group of the frame bundle 
fills out the whole of $O(d)$, there are no singlets in the decomposition of the internal spinors under the structure group of the spin bundle. However,
if the structure group only fills out part of $O(d)$ then singlets may exist.

To illustrate this consider dimensional reduction on a  
Calabi-Yau threefold, which is a six dimensional manifold with $SU(3)$
structure. In fact Calabi-Yau manifolds also have $SU(3)$ holonomy but this additional constraint is not
important here as it relates to the amount of supersymmetry preserved by the four dimensional 
Minkowski space vacuum
rather than the amount associated with the theory. Since the structure group of the frame bundle is
$SU(3)$ the structure group of the associated spin bundle is also
$SU(3)\subset Spin(6)$ where $Spin(6)$ is the double cover of $SO(6)$. 
A spinor of $Spin(6)$
is in the fundamental of $SU(4)$ and so it is easy to see, if one imagines the internal spinor as a four
component column vector, that the $SU(3)$ structure will only leave one of the four independent
components of this spinor representation invariant. More formally the $\bf{4}$ of $SU(4)$ decomposes 
under 
$SU(4) \supset SU(3) \times U(1)$ as $\bf{4} = (\bf{1},3) + (\bf{3},-1)$ where the first numbers in the 
brackets 
are the dimension of the relevant $SU(3)$ representations and the second numbers are the $U(1)$ 
charges. Thus we obtain one $SU(3)$ singlet under this decomposition.
Therefore, standard compactification on a Calabi-Yau threefold (or indeed any 
six dimensional manifold of SU(3) structure where the truncation used has been appropriately defined) 
preserves one quarter of the supersymmetry of the higher dimensional theory.

More generally, this condition for preservation of a certain number of supersymmetries in type II 
reductions would involve 
products of G-structures (see for example \cite{Grana:2005ny}).  This is because the two higher 
dimensional supersymmetry parameters, $\e_{\pm}$,
can be decomposed using two different internal spinors $\chi_{\pm}$ which are associated to different 
G-structures. For example, if one follows this rule, continuing to examine the conditions for $N=2$ 
supersymmetry in a reduction to four dimensions, the relevant condition on the internal manifold is 
that it should have $SU(3)\times SU(3)$ structure. The two internal spinors, $\chi_{\pm}$, which 
we require to be present by demanding $N=2$ supersymmetry, each define an $SU(3)$ structure. 
Locally, these two spinors need not be parallel and so the two $SU(3)$ structures are different
and locally define an $SU(2)$ structure. However,  if the $SU(2)$ structure is globally defined then 
the dimensionally reduced action could instead be written as an $N=4$ theory in four dimensions. 
Therefore, to avoid obtaining a lower dimensional theory that can be written in an $N=4$ manner, the 
two spinors must be somewhere parallel. The $SU(3)$ structure example mentioned above is 
a special case of $SU(3)\times SU(3)$ structure where the two spinors are everywhere parallel 
(see for example \cite{Grana:2005ny}). 

An important point to note here is that although in geometric reductions the higher dimensional 
supersymmetry parameters can be associated to a product of G-structures, this product is defined on 
a single spin bundle (and hence we can consider the intersection of the two structures). 
So even though the spinors 
$\e_{\pm}$ are associated to different G-structures, they are sections of the same spin bundle.
This is because there is only one orthonormal frame bundle with one  associated spin bundle. 
This is one of the features of the discussion which will be modified
in the forthcoming sections when we consider T-folds. 

\subsection{The behaviour of various quantities under T-duality.}
\label{spinortranssubsec}

We wish to propose a modification to the rule for the amount of supersymmetry preserved by 
geometrical reductions, given above, to the case of non-geometric T-fold 
compactifications.
However, before we can describe these modifications we need to 
know how various quantities, in particular the supersymmetry parameters, behave under T-duality 
transformations. \footnote{See appendix A for some comments about local reformulations of the theories
under consideration which are relevant to this and following sections.}.

For simplicity, in the rest of this paper we will consider
the case where, in the lower dimensional theories we are interested in, supermultiplets containing 
fields coming from metric and $B$ field components with a 
single fibre index are truncated. In terms of the massive supergravity presented in the last section 
for example this would 
correspond to the (consistent) truncation where the supermultiplets containing the calligraphic gauge 
fields
(i.e. ${\cal A}_y^{0i}$ and ${\cal A}^{7i}_{\a}$) are set to zero. This truncation of the general case 
provides a vast simplification
in the discussion that will be pursued in this section while still including many of the novel 
mathematical 
(non-) geometric structures that arise in these contexts.
It would be of considerable interest to consider the more general case where these
terms are not zero, however, and this will be pursued by the authors in future work.

Much of what we need to know about the behaviour of the supersymmetry parameters of type II theories 
under
T-duality has been given in the literature \cite{Hassan:1999mm,Hassan:1999bv,Hassan:2000kr}. In this 
work the relevant transformation rules were obtained by examining how the 
supersymmetry variations change under T-duality. 
Using the well known results for how the NS-NS sector transforms, 
such considerations are enough to fix the transformation properties of the spinor parameters.

In this paper we are interested in transformations whose generator, $T$, is not simply that of some 
geometric transformation
such as a rotation. Of the $d(2 d-1)$ generators of the group $SO(d,d)$, $d^2$ correspond 
to general linear coordinate transformations, and $d(d-1)/2$ correspond to B-field shifts. 
As such we shall follow \cite{Hassan:1999mm} in only considering a subgroup of the full $O(d,d)$ 
T-duality group which contains the interesting new cases and which corresponds to 
symmetries of a given theory, rather than dualities which 
swap IIA and IIB. In fact we shall consider an $SO(d) \times SO(d)$ 
subgroup which includes an $SO(d)$ group which is simply the ordinary rotations. We consider this 
slightly 
larger than necessary subgroup as this choice makes the following discussion somewhat clearer.
The generalisation to the $SO(d,d)$ case is
straightforward, and will be discussed in appendix~B.
Our T-duality group elements will take the following form,
\bea
\label{trans}
O = \frac{1}{2} \left( \ba{rr} S+R & S-R \\
S-R & S+R \ea \right) \label{nontrivial}
\eea
where $S$ and $R$ are $(d+b)\times(d+b)$ dimensional matrices which
are related to $SO(d)$ matrices ${\cal S}$ and ${\cal R}$  as follows,
\bea
S = \left( \ba{rr} {\cal S} & \\
& 1_b\ea\right) \qquad
R = \left( \ba{rr} {\cal R} & \\
& 1_b\ea\right)
\eea
Here $d$ is the dimension of the fibre and $b$ is the dimension
of the base.
In this basis our $O(d,d)$ invariant metric is given by the following $2(d+b) \times 2(d+b)$ 
matrix,
\bea
\eta = \left(\ba{rr} 0 & 1 \\  1 & 0 \ea \right)
\eea
and so $O^T \eta O = \eta$ as required. The elements corresponding to ordinary rotations of the 
fibre are those where $S=R$. Note that $O$ is analogous to $\Omega$ as introduced
in section~\ref{theory}. However we have encoded the same information in higher dimensional 
matrices so that we can 
describe the relevant transformations as actions on points living in the entire compact space and 
not just the fibre. As such the dimension of $O$ corresponds to the whole fibre bundle dimension,
while $\Omega$ had dimension $2d$, and thus was associated only to the fibre. 

\vspace{0.2cm}

We now consider the transformation rule for the metric and NS two form
under $O$. As in the 2-dimensional case, the metric and two form on
the fibre bundle (which has overall dimension $d+b$) can be combined into a matrix $M$ which takes 
the same form as before, i.e. 
\bea
\label{newM}
M = \left(\ba{cc} G^{-1} & -G^{-1}B \\ B G^{-1} & G - B G^{-1} B \ea \right)
\eea
where now the blocks are $(d+b)\times (d+b)$ dimensional. Under T-duality $M \rightarrow O
M O^T$. Using the explicit form for $O$ we find the
following transformation property for the inverse metric \cite{Hassan:1999mm,Hassan:1994mq}:
\bea
G^{-1} \rightarrow Q_- G^{-1} Q^T_- = Q_+ G^{-1} Q^T_+
\eea
In the above we have used the following definitions of the $(d+b)$ dimensional matrices $Q$:
\bea
\label{Q}
Q_-= \frac{1}{2} \left[(S+R) + (S-R)(G+B)\right] \\ \nonumber
Q_+= \frac{1}{2} \left[(S+R) - (S-R)(G-B)\right]
\eea
These equations imply two possible different transformations for the
vielbein of the compact space under the transformation \eqref{trans}, either
\bea
\label{viel,Q}
e^M_{\bar{a}} &\rightarrow& \hat{e}^M_{(-)\bar{a}} = Q^M_{-N} e^N_{\bar{a}}  \\ \nonumber
\textnormal{or} \;\;\;\; e^M_{\bar{a}} &\rightarrow& \hat{e}^M_{(+)\bar{a}} = Q^M_{+N} e^N_{\bar{a}} \;.
\eea
Here $M$ is a $(d+b)$-dimensional spacetime index and $\bar{a}$ an orthonormal basis index. 
These two vielbeins are then related by a local Lorentz transformation as follows,
\bea
&\hat{e}&^M_{(+)\bar{b}} = \hat{e}^M_{(-)\bar{a}} \Lambda^{\bar{a}}_{\bar{b}} \;\;\; \textnormal{where} \\
&\Lambda& = e^{-1} Q^{-1}_- Q_+ e
\eea
Physically the situation is as follows. 
 The vielbein $e^M_{\bar{a}}$ can be considered as arising from either the 
left $(+)$ or right $(-)$ moving sector of the string
worldsheet. Under a T-duality transformation it
transforms to either $\hat{e}^M_{(+)\bar{a}}$ or $\hat{e}^M_{(-)\bar{a}}$ depending
on its worldsheet origin. These two local Lorentz frames are
twisted relative to each other by an amount $\Lambda$.
This generalises the statement that a single T-duality on a flat background is 
just parity on one of the world sheet sectors 
to the case of non-trivial backgrounds and more complicated transformations. 

We see then from equation \eqref{viel,Q} that the two
 vielbeins transform differently, and both transformations are
 nonlinear. 
Thus from the point of view of discussing vielbeins on a T-fold we are left with the following 
picture. We define the left and right moving vielbeins separately. The transition functions seen by 
these vielbeins are a nonlinear 
realisation of $SO(d)$ transformations. Thus there are objects which 
resemble an orthonormal frame bundle. The differences to the geometric case lie in the fact that there 
are two `orthonormal frame bundles' with (in general) different transition functions - one for the 
left and one for the right moving sectors. In addition the transition functions on these objects act 
non-linearly. In the limit where we take $S=R$ (which corresponds to
 ordinary rotations) we find that $Q_-=Q_+=S$ and the frame bundles reduce to the usual 
single orthonormal frame bundle with linearly acting transition functions.

\vspace{0.5cm}

Given these transformations
for the orthonormal frames, how do the supersymmetry parameters transform under the T-duality element 
\eqref{trans}? 
We denote the ten dimensional supersymmetry transformation parameters
of our type II theories by $\epsilon_\pm$. We have 
followed \cite{Hassan:1999mm} here in that the subscripts '$\pm$' denote the
worldsheet sector to which the spinor is associated. We 
choose $\epsilon_-$ to have positive chirality in both type II
theories which then fixes the chirality of $\epsilon_+$ in both cases.

We define the spinor representation associated with $\Lambda$ in the usual manner:
\bea
{\Sigma}_{\textnormal{relative}}^{-1} \Gamma^a {\Sigma}_{\textnormal{relative}} = \Lambda^a_{\;b} 
\Gamma^b
\eea
The transformation rules for the supersymmetry parameters have then been found in the paper by 
Hassan, \cite{Hassan:1999mm}, to be,
\bea
\label{spinortrans}
\epsilon_- &\rightarrow& {\Sigma}_{\textnormal{total}}\epsilon_- \\
\epsilon_+ &\rightarrow& {\Sigma}_{\textnormal{total}}{\Sigma}_{\textnormal{relative}}\epsilon_+ 
\eea
where the explicit expression for ${\Sigma}_{\textnormal{relative}}$ is given by,
\bea
{\Sigma_{\textnormal{relative}}} = 2^{-\frac{d}{2}}
\sqrt{\frac{\det({\cal Q}_{-} +
    {\cal Q}_{+})}{\det{{\cal Q}_{-}}}} \left\{ 1 + \sum_{p=1}^{[d/2]}
\frac{(-1)^p}{p!\ 2^p} {\mathcal A}^{i_1 i_2} \dots{\mathcal
  A}^{i_{2p-1} i_{2p}}\Gamma_{i_1i_2\dots i_{2p-1} i_{2p}}   \right\} \;. \label{Srelative}
\eea
where ${\cal Q}_{\pm}$ are the $d\times d$ blocks of
the matrices $Q_{\pm}$, which are associated to the  fibre, and the quantities ${\mathcal A}^{ij}$ are 
components of the following matrix: 
\bea
{\mathcal A}^{ij} = \left[ (1_{d} - {\cal S}^{-1} {\cal R})^{-1}(1_d +
  {\cal S}^{-1} {\cal R}) + {\cal B}\right]^{-1}_{ij}
\eea
where the $\{ij\}$ indices are raised by matrix inversion, and
${\cal B}$ is the $B$-field on the $d$
dimensional fibre. 

Now the work by
Hassan is concerned solely with local considerations, and therefore
the overall rotation on the spinors, ${\Sigma}_{\textnormal{total}}$,
can be set to the identity using the local Lorentz symmetry. This
makes the spinor $\e_-$ invariant under the T-duality transformations.
However, in our case we wish to use the spinor transformations as 
transition functions. Therefore, we are no longer free to
make the choice ${\Sigma}_{\textnormal{total}} = 1$. We must 
specify the total action of the transformations on the spinors up to a
conjugacy class, as this is the degree to which the transition functions are
defined given the presence of local Lorentz symmetry. This should be compared with the similar situation
which exists in the geometric case. There ${\Sigma}_{\textnormal{relative}} = 1$, and 
${\Sigma}_{\textnormal{total}}$ implements the rotation on the spinors associated to the rotation of 
the underlying frame.

There are a number 
of ways in which we can isolate the overall rotation on the spinors. The 
structure of the `orthonormal frame bundles' discussed above makes the correct choice relatively 
clear - one 
should pick the overall rotation so that there is a correspondence between the action on the 
orthonormal frames and the action on the spinors which preserves group
structure. In other words the left and
right moving spin bundles should be `associated to' the left and right moving frame bundles.
One consequence of this link is that, due to the nonlinear and $B$-dependent nature of the transition 
functions on the frame bundles, the transition functions on the spin bundles will in general 
depend on $G$ and $B$ as well as $S$ and $R$. There are a number of
symmetries we expect the spinor transition functions to obey, which
follow from the association with the orthonormal frame
bundles. Firstly, in the limit of flat space with no $B$ field
(i.e. $G+B= 1\!\!1 $), $Q_+$ and $Q_-$ are simply the linearly acting rotations $R$
and $S$ respectively. Therefore, we expect the
transformations on the spinors to be simply given by the spinor
representations of $R$ and $S$ in that case. Secondly, we notice that for a general
background there is a symmetry $Q_+\leftrightarrow Q_-$ if we take $B \rightarrow -B$ and
exchange $S\leftrightarrow R$. This symmetry exchanges the left and
right moving world sheet sectors and so the two frame bundles. Therefore, we expect this same symmetry
to be present in the spinor transition functions; in particular it
should exchange the 
the transition functions for $\e_+$ and $\e_-$.

\vspace{0.2cm}

In practical terms, how do we actually construct the overall rotation? The simplest method is the 
following. Consider a background that is 
a $T^d$ fibration over an $S^1$ base with some monodromy associated with the non-contractible 
loop of the base. We take the monodromy to lie in the non-trivial
subgroup $SO(d) \times SO(d)$ that we have been
discussing. As we traverse the $S^1$, starting at $y=0$ and ending at
$y=2\pi$, the orthonormal frames $e_{\pm}$ undergo field-dependent
transformations implemented by $Q_{\pm}$. Due to the monodromy in the
background these frames do not in general come back to themselves at
$y=2\pi$, but are related to the original frames by some
T-duality transformation. Now consider parallel transporting the
spinors $\e_{\pm}$ around paths where all of the coordinates are constant except for $y$. 
The left moving spinor is transported using the spin connection constructed from the left moving 
vielbein, $e_+$, and similarly for the right moving sector. We find that when
we compare the two spinors with themselves
at $y=0=2\pi$ they have undergone a rotation. This rotation is defined up to a conjugacy class due 
to the 
local Lorentz invariance on the coordinate patch. We get different rotations on the left and right 
moving 
spinors and these have precisely the properties that we require for those associated to the 
monodromy element we have picked -
they obey all of the symmetries mentioned above (this is guaranteed because our parallel transport 
operators involve spin
connections derived from the two vielbeins $e_{\pm}$, and the
vielbeins satisfy these properties) and, in particular, give
the correct relative rotation, ${\Sigma}_{\textnormal{relative}}$,
given in \eqref{Srelative}, which was derived by Hassan using another
method.

By carrying out this construction 
 for an arbitrary monodromy within our $SO(d)\times SO(d)$ subgroup we 
can find all the previously unknown 
transition functions on the supersymmetry parameters. These can then be
 used to construct the spin
bundles associated to a given T-fold. We emphasise that the construction we have outlined here 
need not be based upon the T-fold we are finally interested in. It constitutes 
instead merely a trick for finding transition functions which, once we have them, can be used in any 
appropriate
T-fold context we desire. 

\subsubsection*{Example.}

To give a concrete illustration of the above discussion we will now
present the details for a  simple example. We shall 
examine the case where 
we have a $T^2$ fibre on an $S^1$ base with a $U(1)\times U(1)$ monodromy in a
`nontrivial' subgroup of $SO(2,2,Z)$.

First, we label the
coordinates on the $T^2$ fibre by $z^i$, $i=1,2$, and the coordinate
on the base $S^1$ by $y \in [0, 2\pi]$. We will also use the combined
coordinate $z^A$, $A=1,2,3$, where $z^3\equiv y$. In these
coordinates, the metric on the fibre at $y=0$ is given by the usual torus metric:
\bea
{\cal G}_0 = \frac{\rho_2}{\tau_2} \left(\ba{cc} 1&\tau_1\\ \tau_1 & |\tau|^2\ea\right) 
\eea
where $\tau= \tau_1 + i \tau_2$ is the complex structure modulus, and
$\rho_2$ is the volume modulus. It is convenient to combine the volume modulus
into a complex field $\rho= \rho_1 + i \rho_2$, where
$\rho_1$ is related to the B-field at $y=0$ as follows,
\bea
{\cal B}_0 = \left(\ba{cc} 0 & \rho_1\\
-\rho_1&0\ea\right) \label{Bfield}
\eea
We recall that we are considering the truncation where there are no off-diagonal terms
in the metric or
$B$-field between the fibre and the base, i.e. the components
which have one base and one fibre index are set to zero. Therefore, the metric on the fibre bundle
at $y=0$ is given by
\bea
ds^2 = \frac{\rho_2}{\tau_2}\big{|}dz^1 + \tau dz^2\big{|}^2 +
e^{2\alpha}dy^2 \label{bundlemetric}
\eea  
where $\a$ is a function of the lower-dimensional space.
From the form of the metric above, a natural vielbein and 
inverse vielbein to take at $y=0$ are
\bea
e^{\bar{a}}_A =\tau_2^{-1/2}\left(\ba{ccc} \rho_2^{1/2} &0&0\\ \rho_2^{1/2}\tau_1
& \rho_2^{1/2}\tau_2&0\\
0&0&e^{\a}\tau_2^{1/2} \ea\right) \qquad e^A_{\bar{a}} =
\tau_2^{-1/2}\left(\ba{ccc} \rho_2^{-1/2}\tau_2 & 0&0\\ -\rho_2^{-1/2}\tau_1&
\rho_2^{-1/2}&0\\
0&0&e^{-\a} \tau_2^{1/2}\ea\right) \label{vielbein}
\eea
Now the torus fibre experiences a monodromy around the base
$S^1$. To illustrate our method we choose this mondromy (and so the transition function action we are 
calculating) to take the form 
\bea
O = \left( \ba{cccccc} \cos(a) & 0 & 0 &0& -\sin(a)&0\\
0& \cos(a)&0& \sin(a) & 0&0\\
0&0&1&0&0&0\\
0& -\sin(a)&0& \cos(a)&0&0\\
\sin(a) & 0 &0& 0 &\cos(a)&0\\
0&0&0&0&0&1\ea\right) \label{monodromy}
\eea
which corresponds to the following rotation matrices ${\cal S}$ and
${\cal R}$ (see \eqref{nontrivial}),
\bea
{\cal S} = {\cal R}^T = \left(\ba{cc} \cos(a) & -\sin(a)\\ \sin(a)& \cos(a)\ea \right)
\eea
If we compare with Ref.~\cite{Dabholkar:2002sy} this monodromy lies in
the elliptic conjugacy class of $SL(2)_{\rho}$. In a stringy application we should take $a=\pi$
in order to obtain a non-trivial monodromy within $SL(2,Z)_{\rho}$.
We now require a $y$-dependent $O$ which
  implements this monodromy as we go from $y=0$ to $y=2\pi$. The obvious choice is to replace
  $a\rightarrow a y/2\pi$ in the above expressions. However, it is
  more convenient to choose the following simpler expressions for the base
  space dependent matrices,
\bea
{\cal S}(x(y))\equiv {\cal S}(x) = \frac{1}{\sqrt{1+x^2}} \left(\ba{cc} 1 & - x\\ x &
1\ea\right),\qquad {\cal R}(x) ={\cal S}(x)^T 
\eea
where $x(y) = \tan(ay/2\pi)$ and so $x=0$ and $x=\tan(a)$ are the endpoints
of the loop in the base space (we assume $a \le \pi /2$, so that the
parameterisation makes sense). Note that using $x$ is simply a
coordinate choice for the base, and the final answers will not depend on
this choice (we will see this explicitly from our expressions later
on). These $x$-dependent matrices allow us to
calculate the fibre metric and $B$-field at arbitrary points on the base
space using \eqref{Mtransf}, 
i.e.
\bea
M(x) = O(x)\ M_0\ O(x)^T
\eea
where $M_0$ is the matrix given in \eqref{newM} constructed from the
fibre bundle metric \eqref{bundlemetric} and $B$-field \eqref{Bfield}. Then using all of these $x$-dependent quantities we can
calculate the matrices $Q_{+}(x)$ and $Q_{-}(x)$ which describe the
transformations of the left and right moving frames as a function of the
$S^1$ base's coordinate:
\bea
{e_{(\pm)}}^A_{\bar{a}}(x) = {{Q_{\pm}}^A}_B(x) e^B_{\bar{a}}  \label{basedepviel}
\eea
where $e^B_{\bar{a}}$ is the inverse vielbein at $y=0$ given in
\eqref{vielbein}, and ${Q_{\pm}}(x)$ are constructed from ${\cal
  S}(x)$, ${\cal R}(x)$, and $M(x)$.

We now use the expressions for $e_{(\pm)}(x)$ to calculate the
components of the two associated three dimensional spin connections. In fact, we do not
need to determine all of the components of the connections but simply those which appear in the expressions governing 
parallel transport of the spinors around the relevant closed loops. 
Now the spinors $\e_{\pm}$, which arise from the left and right
moving sectors on the world-sheet, split up into  external
7-dimensional spinors, $\eta_\pm$, and internal 3-dimensional spinors,
$\c_{\pm}$. Since we are interested in the internal space, and because we have made the simplification mentioned at 
the beginning of section IIA, we will
only need to deal with the spinors
$\c_{\pm}$ from now on. The  equations for parallel transport around a path where the fibre coordinates are constant 
are
\be
\nabla^{(\pm)}_x \c_{\pm} = \partial_x \c_{\pm} + \frac{1}{4} {\omega_{(\pm)}}_x^{\bar{a}\bar{b}} \Gamma_{\bar{a}\bar{b}} \c_{\pm}
= 0
\ee
where summation
is assumed over the orthonormal indices $\bar{a},\bar{b}$. 
 It can easily be shown that the expression for the spin connection
 components is
\bea
{\omega_{(\pm)}}^{{\bar{a}}{\bar{b}}}_x =
\sum_A {e_{(\pm)}}^{A[{\bar{a}}}\ \partial_x {e_{(\pm)}}^{\bar{b}]}_A 
\eea
 Using the expressions \eqref{basedepviel} for the left and right moving vielbeins as
 functions of $x$ this
 becomes
\bea
{\omega_{(\pm)}}^{{\bar{a}}{\bar{b}}}_x = - {e}^{[{\bar{a}}}_A
  \left( \partial_x {(Q_{(\pm)}^{-1}(x))^A}_B {(Q_{(\pm)}(x))^B}_C \right) {e}^{{\bar{b}}]C}
\eea
with summation understood over $A, B, C$. Notice that the spin
connection is now written in terms of the original vielbein at $x=0$. 
Using this equation we find that the only non-vanishing spin
connection components are
\be
{\omega_{(-)}}_x^{\bar{1}\bar{2}} = - {\omega_{(+)}}_x^{\bar{1}\bar{2}}= -\frac{2 x^2( x^2 - 1)|\rho|^2 + 8 x^3 \rho_1 - x^4 + 4
  x^2 +1}{|1+ x \rho|^2 (4x^2 |\rho|^2 - 4x (x^2-1)\rho_1 + x^4 - 2x^2 +1)}  
\ee
The simple relation between $\omega_{+}$ and $\omega_{-}$ is due to
the relationship ${\cal S}={\cal R}^T$ for this monodromy. Note also that this spin
connection component does not depend on the complex structure modulus
of the torus. This is because the monodromy is in the $SL(2)_{\rho}$
subgroup of $SO(2,2)$.
We now have the required information to explicitly parallel transport
the spinors around our paths with $z^1=z^2=\textnormal{ constant}$. The resulting
spinors at $x=\tan(a)$ are given by
\bea
\c_{\pm} =
\exp\left(-\frac{1}{2}\int_0^{\tan{a}}{\omega_{(\pm)}}_x^{\bar{1}\bar{2}}
\Gamma_{\bar{1}\bar{2}}\ dx \right) \c_{\pm}^0 \label{totalrot}
\eea
where $\c_{\pm}^0$ are the spinors at $x=0$.
Note that from the above expression it is clear, given that the spin connection contains one $x$ derivative, that we
could have chosen any parameterisation for the closed loop and obtained the
same answer. From our explicit expression for ${\omega_{(\pm)}}_x^{\bar{1}\bar{2}}$ we obtain
\begin{eqnarray}
&&-\frac{1}{2} \int_0^{\tan(a)} {\omega_{(\pm)}}_x^{\bar{1}\bar{2}}\Gamma_{\bar{1}\bar{2}}\ dx = \nonumber\\
&&\qquad \pm \frac{1}{2}\left\{ - \arctan\left(\frac{\rho_1}{\rho_2}\right) +
\arctan\left(\frac{|\rho|^2\tan(a) + \rho_1}{\rho_2}\right) +
\arctan\left(\frac{2\rho_2 \tan(a)}{\tan^2(a) - 1-
  2\tan(a)\rho_1}\right)\right\} \left(\ba{cc} 0&-1\\ 1& 0\ea \right) \label{exponentiatedthing}
\end{eqnarray}
where we have chosen the following realisation for the $\Gamma$
matrices: $\Gamma_{\bar{1}} = \sigma_1$, $\Gamma_{\bar{2}} = \sigma_3$, where
$\sigma_i$ are the usual Pauli matrices. The exponential of this
matrix then gives the explicit matrix which takes $\c_\pm^0$ to
$\c_\pm^{2\pi}$ for this
monodromy. Since we therefore now have the action of the monodromy on both kinds of spinor we have also
determined the overall rotation we have been looking for. 
Clearly the matrix multiplying $\c_+^0$ is the inverse
of that multiplying $\c_-^0$. This is simply a consequence of the choice of monodromy we have made in this example. 
Despite this simple relation, it should be noted that formally the two transition functions associated to $\c_{\pm}$ live in the structure 
groups of two different spinor bundles.

Note that if we take the flat space limit at $y=0$, i.e. take ${\cal G}_0 +
{\cal B}_0 = 1\!\!1$ (which corresponds to taking $\rho=\tau = i$, although
note that the $\tau$ modulus does not actually appear, so its
value doesn't make a difference here), then the
factor multiplying the constant matrix in \eqref{exponentiatedthing}
becomes $\mp a/2$. This is $\mp 1/2$ times the angle which appears in the monodromy matrix. This
makes sense because in the flat space limit $Q_-= S$, $Q_+ = R$ and so
the spinors transform via the spin representation of these
rotation matrices. From \eqref{exponentiatedthing} we see that
another expected feature of these transformations also holds, namely that taking $a\rightarrow
-a$ and $\rho_1\rightarrow -\rho_1$ exchanges the transformation matrices
for $\c_-$ and $\c_+$.

\vspace{0.2cm}

As a brief aside before we move on to discuss the preservation of supersymmetry, we note that in the 
2-torus case, the matrix multiplying
$\c_{\pm}^0$ in \eqref{totalrot} can be written in a nicer form as the spin representation of the 
following matrix 
\bea
A_{\pm} = \left(\ba{cc} \frac{1}{\sqrt{\det({\cal Q}_{\pm})}}\tilde{e}^{-1}
{\cal Q}_{\pm} \tilde{e} &0\\
0& 1\ea\right)
\eea
where ${\cal Q}_\pm$ is evaluated at $x=\tan(a)$ and $\tilde{e}$ is
the $2\times 2$ part of the vielbein $e$. It turns out that in the
$O(2,2)$ case, these matrices are orthogonal. In other words,
\bea
{\Sigma} ( A_{\pm} ) =\exp\left(-\frac{1}{2}\int_0^{\tan{a}}{\omega_{(\pm)}}_x^{\bar{1}\bar{2}}
\Gamma_{\bar{1}\bar{2}}\ dx \right) 
\eea
where $\Sigma$ denotes the spin representation,
obtained in the usual way. Similar comments do not hold
in the case of higher dimensional fibres however. 

\vspace{0.2cm}

In any case we now set aside our example and return to the general discussion to show how to determine 
the amount of supersymmetry preserved by appropriately defined T-fold reductions.

\subsection{`G-structures' and supersymmetry in the non-geometric case.}\label{susy}

We are now going to describe a proposal for a simple rule for determining the amount of
supersymmetry preserved by T-fold reductions of the form of a $T^d$
bundle over $S^1$ when a sensible truncation is defined in the dimensional 
reduction process.
Our discussion generalises in an obvious way to more complicated T-fold reductions.
As discussed earlier, we are 
interested in describing the supersymmetry preserved by the
compactified {\it theory} rather than by any particular lower
dimensional vacuum. Therefore, we will follow the analogue of the G-structures story 
that we outlined at the start of this section. In addition we repeat that we are considering the 
case where supermultiplets with fields originating from higher dimensional metric and $B$ field 
components with a single fibre index have been truncated and that we are not
considering the 'non-minimal' cases discussed in the appendix A.

\vspace{0.2cm}

The rule we propose for the amount of supersymmetry preserved by T-fold reduction in fact follows along 
fairly similar lines to the geometric case if the situation is phrased
in terms of structure groups of spin bundles. 
As in the geometric case we decompose the spinor 
parameters into sums of internal and external pieces. We then consider how the internal pieces of these 
spinors transform under the structure groups of the relevant spin
bundles (of which there are now two). We propose that the number of singlets in the
decomposition of each
internal spinor into representations of the relevant structure group then determines how many lower 
dimensional supersymmetries are associated to it in the case of an appropriately defined truncation.

However, there are crucial differences between the non-geometric and
geometric cases. One obvious difference is that on T-folds 
the transition functions on the spinor supersymmetry parameters are
generally $B$ and $G$ dependent, as we have seen explicitly in the
previous section. 
This 
point deserves a little further discussion. At a first glance one might naively assume that this fact,
coupled with the above comments, means that the amount of supersymmetry of the lower dimensional {\it 
theory} is a moduli dependent quantity! This clearly can not be the case. In fact since we are talking 
about the theory rather than about any particular lower dimensional vacuum (and so moduli values) the 
correct amount of supersymmetry is the minimum that is obtained when arbitrary values of the moduli are 
allowed. 

We could however create a different lower dimensional theory if a consistent truncation were 
available where some of the lower dimensional fields were set to
certain constant values (indeed we have already done something very similar with the fields we have 
already truncated). This could conceivably result in a truncated theory with 
more supersymmetry than its parent theory. Such phenomena are well known within supersymmetric 
compactifications and it should not come as a surprise that examples could also appear in this context.

The fact that we have two different spin bundles with different transition functions on them is
major difference between the case at hand and that of geometric compactification. For example, the 
existence of two globally well defined, nowhere parallel internal spinors in the geometric case 
would imply the existence of 4 supersymmetries in the dimensionally reduced theory if the 
usual rule where to hold (i.e. two from each of $\e_+$ and $\e_-$). However, in the T-fold case, the same need not 
necessarily be true. The two internal spinors could be sections of
different spin bundles. This would lead to us 
obtaining one lower dimensional supersymmetry from each higher dimensional spinor parameter - 
giving us a total of two rather than four.

Let us examine now the examples of section II where we performed a
Scherk-Schwarz reduction and truncation. Do these examples obey the
rule we have proposed for the number of supersymmetries in the lower
dimensional theory?
In fact, this is a relatively 
easy question to answer.

We have, by the nature of our reduction ansatz \eqref{ansatz}, 
truncated the higher dimensional theory to 
consider only field configurations of the form
\bea
\psi(y) = g(y)[\psi_0]
\eea
Which supersymmetries are compatible with this truncation? Clearly from our earlier discussion only 
those supersymmetry transformations which, when applied to field configurations of this form give 
back another field configuration also in this form are compatible. 
Applying a supersymmetry transformation to a field configuration of this form we find the following,
\bea
(1+\delta_{\e }) \psi(y) &=& (1+\delta_{\e }) g(y)[\psi_0] \\
&=& g(y)[(1+\delta_{\Sigma^{-1} \e }) \psi_0]
\eea
Thus only if $(1+\delta_{\Sigma^{-1} \e }) \psi_0=\psi_0'$, where $\psi_0'$
is independent of $y$, do we get back to a field configuration of
the same form. Given $\Sigma$'s dependence on $y$ in our examples this means that only 
supersymmetries of the form $\e = \Sigma \e_0$, where $\e_0$ is not dependent on the internal 
coordinates, are compatible with the truncation. However, such spinors are not globally well defined unless $\Sigma \e_0 = \e_0$. In other cases the spinor $\e$ will 
be double valued at $y=0,2 \pi$. Thus as our rule states, the
supersymmetries which are preserved in Scherk-Schwarz reduction arise
from  internal spinors which are singlets under the monodromy (and so the structure
group).

\vspace{0.2cm}
 
Clearly there is a lot of interesting (non)-geometrical structure in the above. 
In addition we have now answered the question we set out to address: we know how to calculate 
the amount of supersymmetry preserved by certain T-fold reductions. 

\subsubsection*{Some Examples.}

Let us illustrate our discussion by working out the amount of supersymmetry associated with the 
appropriate 
truncations (dropping the supermultiplets containing the calligraphic fields of section II) of the 
massive supergravity theories discussed earlier in the paper. We therefore consider the case where 
we have a $T^d$ fibration over an $S^1$ base. The structure group of the spin bundles is simply 
given by the monodromy $e^T$ which is introduced into the background. 
Therefore, in all of our cases the monodromy simply constitutes a single operator which then 
lies in a $U(1)$ 
subgroup of $O(d,d)$. One can then use the discussion of this section to determine, for any given 
monodromy, how this $U(1)$ acts on the left and right moving spinors.
For the case of a $T^2$ fibre the ${\bf 16}$ of $SO(9,1)$ breaks up into $({\bf 8},{\bf 2})$ 
representations of $SO(6,1) \times
SO(3)$. Now clearly because the spinor of $SO(3)$ is in the fundamental of 
$SU(2)$, a $U(1)$ structure group will leave none of the degrees of
freedom of the  ${\bf 2}$ invariant. For general 
values of $B$ and $G$ the monodromy will result in a non-trivial
structure group for both spin bundles, as can be seen from our example of the previous subsection.
Therefore in general 
we will be left with none of the higher dimensional supersymmetries being present 
in the lower dimensional theory. Thus in the general case the
resulting 7-dimensional theory, as presented in section II, has no supersymmetry. 

However, as
discussed earlier, if one takes a truncation of the lower dimensional
theory then it is possible that this truncated theory could contain more
supersymmetry than its parent. Let us consider the particular example presented 
in section~\ref{spinortranssubsec} where we chose a certain
monodromy for the $T^2$ fibre \eqref{monodromy} and see if this occurs in this case. By considering the
scalar potential \eqref{scalarpotential} associated to this monodromy, one finds that a consistent 
truncation can be made by setting $\rho=i$, as this minimises the potential for the relevant moduli. 
However in this case, even after such a truncation, we see from our expressions at the end of 
section III that the monodromy still acts on both types of spinor in a non-trivial way. Thus for this 
particular choice of monodromy the truncated theory also has no supersymmetry and the phenomenon of 
enhancement of symmetry on truncation does not occur.

For the case of a $T^3$ fibre, the story for the general case
is slightly different. 
The ${\bf 16}$ of $SO(9,1)$ breaks up into 
$({\bf 4},{\bf 2})+({\bf 4}',{\bf 2}')$ representations of $SO(5,1) \times
SO(4)$. Now the two Weyl spinors of $SO(4)$ that arise here transform under different $SU(2)$ 
subgroups and so one of these 
will remain invariant if we arrange matters so that the $U(1)$ is a subgroup of the other $SU(2)$. 
The analysis of the other $SU(2)$ will follow as in the case of the $T^2$ fibre.
Therefore, in general half the supersymmetries associated to $\e_+$ and
$\e_-$ are preserved, and so half of the possible supersymmetries in the six dimensional theory 
are preserved, so we obtain either $(1,1)$ or $(2,0)$ supersymmetry in 6 dimensions. 

As before, a consistent truncation of the lower dimensional theory could result in more supersymmetry 
with different numbers of supersymmetries coming from the two higher dimensional supersymmetry parameters 
for certain choices of monodromy.

Similar analyses can be pursued for all higher dimensional fibres. 
Clearly the arguments described in this section can also easily be extended
to deal with more general T-fold constructions, where the base is not
simply $S^1$.

\section{Conclusions.}
\label{conc}

In this paper we have considered the dimensional reduction of type II supergravity theories
on T-folds. We began by briefly discussing the dimensional reduction of higher dimensional
actions to obtain lower dimensional massive supergravity theories. We then went on to discuss some 
underlying (non-) geometric structures associated with these compactifications in order to 
calculate how much supersymmetry the lower dimensional theories possess.

We showed that associated with these spaces there are two `orthonormal frame bundles', one 
associated with
each of the left and right moving sectors of the string worldsheet. These have non-linearly acting,
field dependent transition functions on them. Associated to these orthonormal frame bundles, in some 
sense, are two spin bundles. Due to the nature of the action of T-duality on the relevant spinors 
these also have field dependent transition functions. By examining the subset of the full field 
dependent representation of $SO(d)$
that these transition functions take in any one case, we were able to identify any spinors that are 
invariant under the structure group of the bundle of which they are sections. 
As in the geometric case this information then told us
how much supersymmetry to expect in an appropriately defined 
dimensionally reduced theory. All of the structure we 
described reduces to the usual structure of a single orthonormal frame and spin bundle etc.
when the structure group is taken to lie within a geometric 
subgroup.
We finally illustrated our discussion by calculating the amount of supersymmetry associated with 
truncations of various 
massive supergravity theories that have been presented in the literature and in the early 
part of this paper. It should be noted that we have been 
concerned in this paper with the amount of supersymmetry associated to the compactified {\it theory} 
rather than any one particular lower dimensional vacuum. 

\vspace{0.2cm}

There is clearly much work still to be done on the investigation of various properties of T-folds. 
One of the 
more interesting directions for future work would be to look for connections between the
structure presented here and mathematical frameworks such as 
generalised complex geometry.

\section{Acknowledgements}

We would like to thank the following people for extremely useful discussions and emails, J. Figueroa-O'Farrill,
S.F. Hassan, V. Jejjala and S. Ross. We would also like to especially thank S. Morris for collaboration during the 
early stages of this project, D. Smith and C. Hull for very many helpful discussions and comments and D. Waldram for very helpful comments on the 
manuscript. J.G. is funded by PPARC and E. H.-J. by EPSRC.


\section*{Appendix A: Coset reformulations.}

In this paper when we have considered  type II supergravity compactified 
on a torus we have used the formulation of this theory with the minimum number of degrees of 
freedom in it. There are various other formulations of these theories. Indeed two such 
reformulations are given in the paper of Maharana and Schwarz \cite{Maharana:1992my}. 
These descriptions of the
theory essentially constitute introducing extra degrees of freedom into the theory as well as 
extra auxiliary gauge symmetries in order to remove them again. This kind of procedure might 
be carried out, for example, in order to linearise the action of the T-duality transformations 
acting on the fields. Locally these reformulations of the theory are all physically equivalent.
One can recover the description of the theory used in this paper from one of these coset 
formulations merely by stipulating a specific gauge choice for the auxiliaries.
Globally, however, the formulations can differ. The presence of the auxiliary gauge symmetries can lead to 
a richer global structure than is present in the minimal version of the theory. One could, for 
example, have Wilson lines in these gauge fields. It would then be impossible to choose a gauge to recover the minimal formulation on every coordinate patch at the same time.

In this paper we only consider the standard formulation and the supersymmetry of compactifications on 
T-folds which are constructed within this framework. It would clearly be of interest to study 
the coset reformulations of this theory in this context as
well. However, this is beyond the 
scope of this paper. It should be pointed out that our results do apply to a subset of 
the configurations that can result from these other formulations. The relevant subset is that where the 
global structure of the auxiliary gauge fields is such that we can make the gauge choice 
to restore the minimal formulation simultaneously on every coordinate patch \footnote{We 
would like to thank Chris Hull for pointing out the importance of these reformulations of 
the theory in this context.}.


\section*{Appendix B: Extension to more general $SO(d,d)$ monodromies}\label{sodd}
In the main body of this paper we have considered monodromies which
lie in a $d(d-1)/2$-dimensional subgroup of $SO(d,d)$ and take the
``non-trivial'' form given in \eqref{nontrivial}. However, we can
consider more general elements of $SO(d,d)$
by combining these non-trivial twists with general coordinate transformations and
B-shifts. In the case $d=2$ we can generate a general element of
$SO(2,2)$ by the following product of group elements,
\bea
{O}_{\textnormal{total}} = \Lambda B O
\eea
 where $\Lambda$ is a coordinate transformation matrix, $B$ defines a B-shift and $O$ is a non-trivial
$SO(2)\times SO(2)$ matrix of the form \eqref{nontrivial}. Explicitly, 
the B-shifts take
the following form, 
\[
 B_i = \left( \begin{array}{cc}
1 & 0 \\
b_i& 1\end{array}\right)
\]
 and the matrix $\Lambda$ is
given by 
\[
\Lambda = \left( \begin{array}{cc}
\lambda & 0\\
0 & (\lambda^T)^{-1}\end{array}\right)
\] 
where $\lambda$ is the $2 \times 2$ matrix
associated to the coordinate
transformation.
We
now consider the action of the element ${O}_{\textnormal{total}}$ on the spinors $\epsilon_+$
and $\epsilon_{-}$. Firstly, we note that B-shifts have no effect on
either spinor parameters, but the value of $B$ does enter the 
matrices ${\Sigma}^{(\pm)}$, which are the spinor transformation
matrices associated to the non-trivial element $O$. So the effective action on the spinors
$\epsilon_{\pm}$  is 
\bea
\epsilon_+ &\rightarrow& {\cal L}\ {\Sigma}^{(+)} \epsilon_+\\
\epsilon_- &\rightarrow& {\cal L}\ {\Sigma}^{(-)} \epsilon_-
\eea  
where ${\cal L}$ is the spin representation of the coordinate
transformation $\Lambda$ (note that the spinors will only ``see'' the
rotation part of the matrix $\Lambda$). 
For $d\geq 3$ the situation is slightly different as we can generate all
elements of $SO(d,d)$ from B-shifts, coordinate rotations (as opposed
to general coordinate transformations) and
non-trivial twists. This can be seen by considering the Lie algebra
elements associated to each of the transformations, and by calculating
their commutators. One finds that a general element of
$SO(d,d)$, $d\geq 3$, can be written as 
\bea 
{O}_{\textnormal{total}} = L O_3 B_3 O_2 B_2 O_1 
\eea
where $L$ is a coordinate rotation. 
The corresponding spinor transformations are then
\bea
\epsilon_+ &\rightarrow& {\cal L}\ {\Sigma}^{(+)}_3 {\Sigma}^{(+)}_2 {\Sigma}^{(+)}_1 \epsilon_+\\
\epsilon_- &\rightarrow& {\cal L}\ {\Sigma}^{(-)}_3 {\Sigma}^{(-)}_2 {\Sigma}^{(-)}_1 \epsilon_-
\eea
Note that the fact that $B$ changes is important in these
transformations, as a different value of $B$ will enter each ${\Sigma}_i$. With this information we could now tackle a T-fold with
monodromy in the full connected subgroup of $SO(d,d)$.


\end{document}